\documentclass[letter]{aa}

\usepackage{graphicx}

\usepackage{txfonts}

\usepackage{xcolor}
\usepackage[utf8]{inputenc}
\definecolor{xlinkcolor}{cmyk}{1,0.6,0,0}

\usepackage[breaklinks=true, 
    bookmarks=false,         
     pdfnewwindow=true,      
     colorlinks=true,   
     linkcolor=xlinkcolor,     
     citecolor=xlinkcolor,    
     filecolor=xlinkcolor,  
     urlcolor=xlinkcolor,      
final=true
]{hyperref}

\begin{document}

   \title{Discovery of a galaxy associated with the HI cloud FAST J0139+4328}

   \subtitle{}

\titlerunning{Optical counterpart of FAST J0139+4328}

   \author{Ana Mitra{\v s}inovi{\' c}\inst{1}
            \and
          Marko Grozdanovi{\' c}\inst{1}
            \and
          Ana Lalovi{\' c}\inst{1}
            \and
          Milena Jovanovi{\' c}\inst{1}
            \and
          Michal B{\' i}lek\inst{1}
            \and
          Nata{\v s}a Pavlov\inst{2}
            \and
          Alexei V. Moiseev\inst{3} 
                   \and
          Dmitry V. Oparin\inst{3}          
          }

   \institute{$^1$ Astronomical Observatory, Volgina 7, 11060 Belgrade, Serbia\\
              \email{amitrasinovic@aob.rs}\\
   $^2$ Faculty of Mathematics, University of Belgrade, Studentski trg 16, 11158 Belgrade, Serbia\\
   $^3$ Special Astrophysical Observatory, Russian Academy of Sciences, Nizhny Arkhyz 369167, Russia 
     }

   \date{Received ; accepted }

  \abstract
  {The search for ``dark galaxies,'' a key prediction of the lambda cold dark matter, has yielded few viable candidates. Recently, FAST J0139+4328 was reported as the first isolated dark galaxy in the nearby universe, based on a neutral hydrogen (HI) detection and a non-detection in the Pan-STARRS1 survey. To verify the nature of this candidate, we obtained deep optical imaging, using the $1.4\,\mathrm{m}$ \textit{Milanković} and $0.6\,\mathrm{m}$ \textit{Nedeljković} telescopes, and spectroscopic follow-up of the field. We report the unambiguous discovery of a low-surface-brightness (LSB) optical counterpart at the location of the HI cloud. Furthermore, the detection of H$\alpha$ emission via the $6\,\mathrm{m}$ Big Telescope Alt-Azimuthal (BTA) confirms that the stellar system lies at a redshift consistent with the HI source, establishing their physical association. Through detailed photometry and employing color-dependent mass-to-light scaling relations, we derive a total stellar mass of $M_\star = (7.2 \pm 3.7) \times 10^6\, M_{\odot}$, about an order of magnitude higher than the previously estimated upper limit. Using the literature HI mass, this implies a gas-to-stellar mass ratio of $M_{\mathrm{HI}} / M_{\star} = 11.5 \pm 6.4$. Our findings demonstrate that FAST J0139+4328 is not a dark galaxy but an extremely gas-rich LSB dwarf galaxy, whose stellar component was simply below the detection limit of the Pan-STARRS1 survey. This reclassification resolves the status of this prominent dark galaxy candidate and underscores the necessity of deep optical follow-up to classify faint HI-selected systems.}
  

   \maketitle
%

\section{Introduction}

The standard lambda cold dark matter ($\Lambda$CDM) cosmological model, while remarkably successful and widely adopted, still faces challenges, especially at small scales \citep[for a review, see][]{Bullock+BoylanKolchin2017ARA&A..55..343B}. One of these challenges is the so-called ``missing satellite problem'' \citep{Klypin+1999ApJ...522...82K, Moore+1999ApJ...524L..19M}: the discrepancy between the high number of low-mass dark matter (DM) halos predicted by simulations and the far fewer observed dwarf galaxies. The problem is often considered solved by invoking the existence of ``dark galaxies'' -- DM halos with no or very few stars \citep{Simon+Geha2007ApJ...670..313S, Sawala+2016MNRAS.457.1931S, Wetzel+2016ApJ...827L..23W, Simpson+2018MNRAS.478..548S, Buck+2019MNRAS.483.1314B, Sales+2022NatAs...6..897S, Jeon+2025ApJ...988..136J}. These objects may contain neutral hydrogen (HI) but have failed to form significant stellar populations, rendering them invisible to traditional optical surveys. Thus, the most effective way to find such objects is through HI surveys \citep[e.g.,][]{Taylor+2012MNRAS.423..787T, Taylor+2013MNRAS.428..459T, Taylor+2022AJ....164..233T, Koribalski+2020Ap&SS.365..118K, Deg+2022PASA...39...59D, Murugeshan+2024PASA...41...88M, Zhang+2024SCPMA..6719511Z, Kwon+2025ApJS..279...38K}.

The systematic search for starless HI clouds has, however, led to important discoveries about the faint universe. In particular, the ALFALFA survey\footnote{It stands for the Arecibo Legacy Fast ALFA, which was a massive, blind extragalactic survey for HI, carried out using the ALFA (Arecibo L-band Feed Array) receiver on the Arecibo radio telescope.} \citep{Haynes+2018ApJ...861...49H} has been instrumental in populating the low-mass end of the galaxy mass function. The most promising dark galaxy candidates have often been reclassified upon further investigation. Some, including the famous VIRGOHI 21 \citep{Davies+2004MNRAS.349..922D}, were ultimately shown to be tidal debris rather than self-gravitating halos \citep{Haynes+2007ApJ...665L..19H, Duc+Bournaud2008ApJ...673..787D}. In many other cases, what was initially identified as an isolated HI cloud with no optical counterpart was later revealed, through deeper imaging, to be a faint stellar system \citep[e.g.,][]{Matsuoka+2012AJ....144..159M, Janowiecki+2015ApJ...801...96J, Janesh+2017ApJ...837L..16J, Janesh+2019AJ....157..183J, Leisman+2021AJ....162..274L, Du+2024ApJ...964...85D, Jones+2024ApJ...966L..15J, OBeirne+2025PASA...42...87O}. The ALFALFA survey has thus been highly effective in identifying a diverse population of low-surface-brightness (LSB) galaxies. This outcome underscores the critical need for deep optical follow-up to distinguish between truly dark objects and extremely faint (star-forming) LSB systems.

The high sensitivity of the Five-hundred-meter Aperture Spherical radio Telescope (FAST) has recently enabled new, deep searches \citep{Jiang+2019SCPMA..6259502J, Jiang+2020RAA....20...64J}. Using FAST, \citet{Xu+2023ApJ...944L..40X} discovered FAST J0139+4328, an isolated HI cloud with kinematics suggestive of a rotating disk. Based on its isolation and the lack of an optical counterpart in Pan-STARRS1\footnote{The Pan-STARRS1 survey, which stands for Panoramic Survey Telescope and Rapid Response System, is a 3$\pi$ steradian survey with a medium depth in five bands \citep{Chambers+2016arXiv161205560C}.} images, it was classified as the first isolated dark dwarf galaxy in the nearby universe. Although a very promising candidate -- because its isolation makes the tidal origin unlikely -- the ``dark'' classification of FAST J0139+4328 was, however, contingent on the depth of the Pan-STARRS1 survey. This was rightly pointed out by \citet{OBeirne+2025PASA...42...87O}, who emphasized that the surface brightness limit of the survey is just $24\;\mathrm{mag}\;\mathrm{arcsec}^{-2}$ \citep{Sola+2022A&A...662A.124S}, insufficient to detect fainter LSB populations. Hence, deeper imaging is required to definitively rule out a faint stellar counterpart. With sufficiently long integrations, small- and medium-aperture telescopes are capable of reaching LSB depths ($\sim$28-28.5 mag arcsec$^{-2}$) that can surpass those obtainable with large, professional facilities \citep{dgsat}. Dithering enhances the recovery of LSB structures by improving sky modeling and mitigating flat-field and charge-coupled device (CCD) systematics, which dominate photometric errors in deep imaging. Observations with the telescopes at the Astronomical Station Vidojevica (Serbia) have proven their capability for detecting and analyzing LSB features \citep[e.g.,][]{oliver, Bilek+2020A&A...642L..10B, ebrova}.

In this letter, we present new, deep optical observations, following the procedure of \citet{Bilek+2020A&A...642L..10B}. Our observations reveal the faint stellar counterpart to FAST J0139+4328. We reclassify the object as a gas-rich LSB galaxy and, adopting the distance from the HI cloud \citep[i.e., from][]{Xu+2023ApJ...944L..40X}, present its photometric properties and a preliminary stellar mass estimate. We also confirm the physical association between these two objects by detecting an ionized gas at the same redshift as the HI cloud.

\section{Observations and data reduction}\label{sec:obs}
Observations were carried out using the $1.4\,\mathrm{m}$ \textit{Milanković} telescope and $0.6\,\mathrm{m}$ \textit{Nedeljković} telescope at the Astronomical Station Vidojevica, operated by the Astronomical Observatory of Belgrade, Serbia. Data were obtained using the Luminance $L$, Johnson $B$, and $V$ photometric bands on the $1.4\,\mathrm{m}$ telescope and the near-infrared $I$ band on the $0.6\,\mathrm{m}$ telescope, with corresponding exposures of $90\,\mathrm{s}$, $180\,\mathrm{s}$, $180\,\mathrm{s}$, and $240\,\mathrm{s}$ during four photometric nights with seeing around $1$ arcsec. The total exposure times are $7\,\mathrm{h}$, $6.35\,\mathrm{h}$, $4.35\,\mathrm{h}$, and $9.66\,\mathrm{h}$ in the $LBVI$ bands, respectively. The detector used on the $1.4\,\mathrm{m}$ telescope was an iKon-L 936 CCD camera with a $2048\times 2048$ pixel array and a plate scale of $0.39$ arcsec per pixel, providing a total field of view of $13.3 \times 13.3$ arc minutes, and an FLI ProLine PL23042 CCD on the $0.6\,\mathrm{m}$ telescope with a $2048\times 2064$ pixel array and a plate scale of $0.512$ arcsec per pixel, providing a total field of view of $17.5 \times 17.5$ arc minutes. A dithering pattern of 300 arcsec was applied for the L filter, 150 arcsec for the B and V filters, and 400 arcsec for the I filter between exposures to improve the accuracy of background estimation while combining the images. Even small uncertainties in sky subtraction or residual background systematics can significantly affect the measurements of central surface brightness and total flux in LSB regions \citep{Trujillo2016}. Although the central brightness is moderately robust, the total flux is highly sensitive to sky errors, emphasizing the need for accurate background modeling \citep{Mihos2019}.

Photometry was calibrated using the Fourth US Naval Observatory CCD Astrograph Catalog \citep[UCAC4,][]{ucac4} selecting bright, unsaturated stars in the $B$-, $V$- and $I$-band images. In total, 62 stars were selected spanning 4 magnitudes in range. Zero points were estimated from linear regression as intercepts (with the slope fixed to unity): 29.08 and 30.24 in the $B$ and $V$ bands, respectively. The $I$-band image was calibrated using relations between the Johnson–Cousins $UBVRI$ and the SDSS $ugriz$ photometric systems, with magnitudes from the UCAC4 catalog. The calibration was calculated from the relations given in \cite{Icalibration}, which can be reduced to the expression $I = 1.245\ i - 0.245\ r - 0.387$. All magnitude zero points were corrected for Galactic extinction following \cite{extinction}: 0.332, 0.251, and 0.138 mag for the $B
VI$ bands, respectively. The $L$-band image was taken specifically to detect the object prior to other (subsequent) observations and was not calibrated, since no further conclusions regarding galaxy physical parameters could be inferred from it. On the other hand, the $(B-V)$ color was used to estimate the stellar mass of the galaxy (Sect. \ref{subsection:stellarmass}) and is plotted in Fig.~\ref{fig:sb}. The $I$-band image was used to measure the half-light radius of the galaxy from the cumulative distribution of the galaxy intensity using Astropy's Photutils Isophote package \citep{astropy2013, astropy2018, photutils2023, ellipse}.

Optical spectroscopic observations were carried out at the $6\,\mathrm{m}$ Big Telescope Alt-Azimuthal (BTA) of the Special Astrophysical Observatory of the Russian Academy of Sciences (SAO RAS) using the multimode focal reducer SCORPIO-2 \citep{scopio2} in long-slit mode. On November 21 and 22, 2025, a total exposure time of $3\,\mathrm{h}$ was collected. The spectrograph slit (with length 6\farcm4 and width $1''$) was placed along the major axis of the galaxy with a position angle $PA=126^\circ$. The atmospheric seeing was about $2.0''$, the spectral range was 3650--7300 \AA, and the spectral resolution was $\sim5$\AA\, with a dispersion of $0.9''$/px and a spatial sampling along the slit of $0.39''$/px. Data reduction was performed using a standard pipeline package running in the Interactive Data Language (IDL) environment, designed for SCORPIO-2 long-slit data \citep{Arshinova2025}. The flux calibration was performed using observations of the spectrophotometric standard star BD+25~4655 on the same night before the galaxy observations.

\section{Results and discussion}\label{sec:results}

\begin{figure}
    \centering
    \includegraphics[width=1\linewidth]{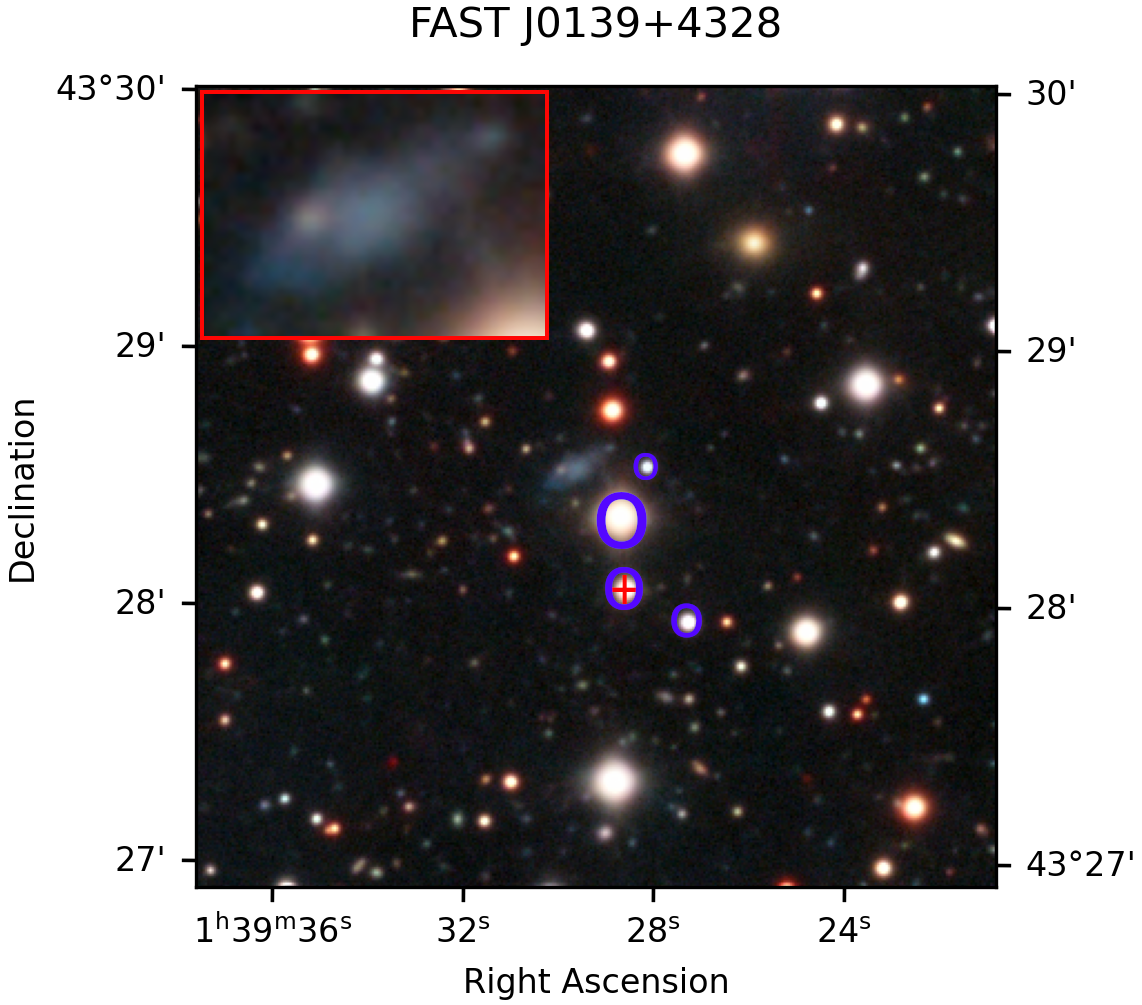}
    \caption{LBVI multiband image with magnified galaxy (inset, upper left corner). The position of the HI cloud is marked with the red plus sign. The circles mark the foreground stars identified using Gaia, corresponding to the same sources highlighted in \citet{Xu+2023ApJ...944L..40X}.}
    \label{fig:LBVI}
\end{figure}

The multicolor $LBVI$ composite (Fig.~\ref{fig:LBVI}) was created to provide a first preview of the appearance and color of the galaxy, while detailed color photometry will be presented in future work. The stacked images in $B$, $V$, and $I$ were combined into a color composite by assigning $B$ to the blue, $V$ to the green, and $I$ to the red channel. Data in the $L$ band were used as a luminance layer to enhance detail and contrast.
The faint blue structure at the center of the image corresponds to the optical counterpart of the HI cloud, which is also magnified and inset. The position of the HI cloud is marked with a red plus sign.

Evidently, unlike previous surveys, our deep imaging reveals a clear optical counterpart. The galaxy is centered at $\alpha (J2000) = 01^\mathrm{h}39^\mathrm{m}29\fs6$, $\delta (J2000) = +43\degr28\arcmin29\farcs8$. This position is offset by $\sim30.3\,\mathrm{arcsec}$ from the HI coordinates reported in \citet{Xu+2023ApJ...944L..40X}. The small displacement is negligible and well within the $\sim 3\,\mathrm{arcmin}$ FAST beam. Such and even larger offsets (but within the beam) are typical when cross-matching large-beam HI detections with optical sources \citep[see][]{Zhang+2024SCPMA..6719511Z}.

\begin{figure}
    \centering
    \includegraphics[width=1.\linewidth,height=1.\linewidth]{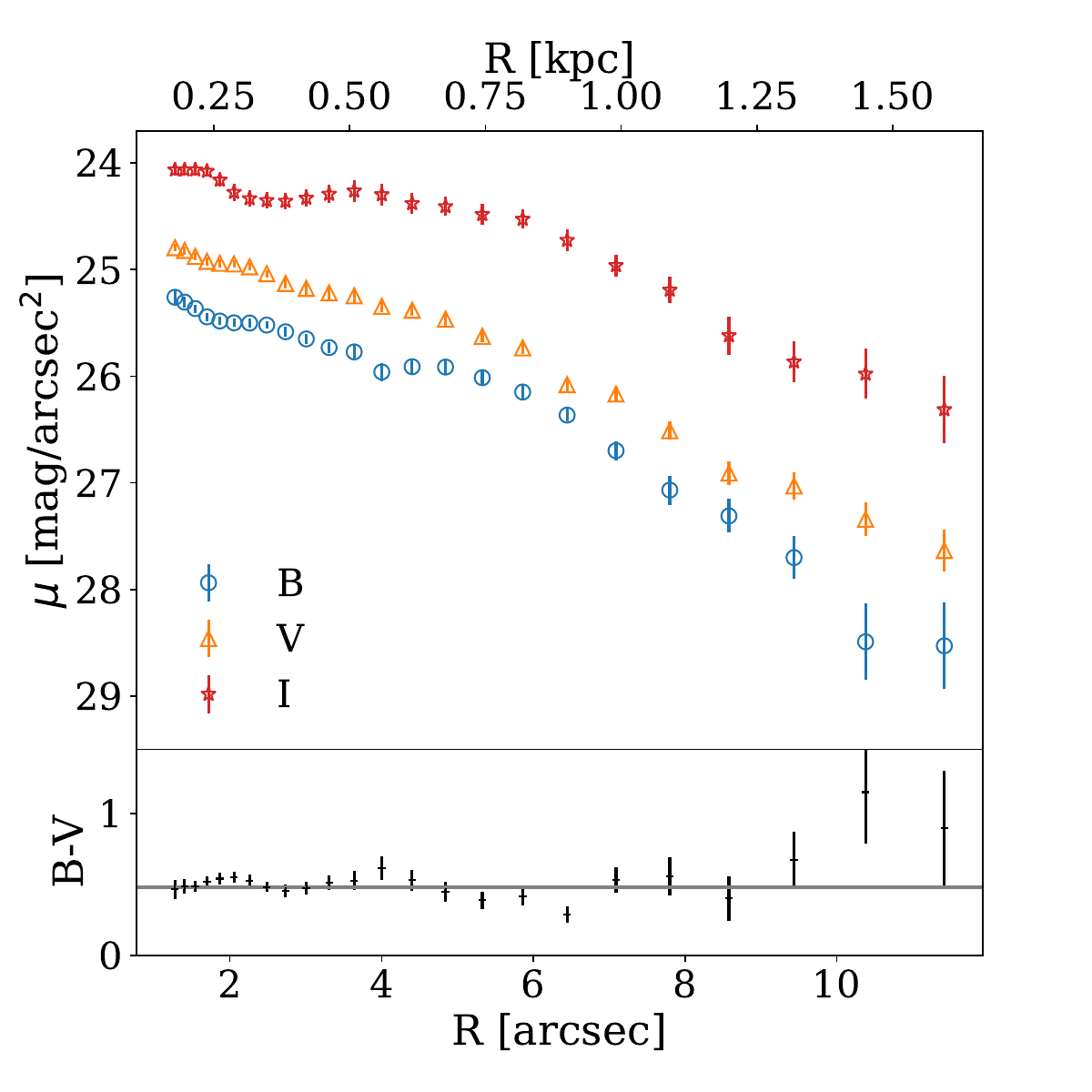}
    \caption{Surface brightness profile of the FAST J0139+4328 galaxy in the $BVI$ bands in the Vega system, corrected for Galactic extinction. In the lower panel, the $B-V$ color is shown with a horizontal gray line indicating its mean value (see Sect.~\ref{subsection:stellarmass}).}
    \label{fig:sb}
\end{figure}

The stellar half-light radius was determined as described in Sect.~\ref{sec:obs}, resulting in $r_{\rm eff}$ = 5.24 arcsec, which corresponds to $r_{\rm eff}$ = 0.73 kpc for the assumed distance of $28.8$ Mpc \citep{Xu+2023ApJ...944L..40X}. The surface brightness distribution of the galaxy is shown in Fig.~\ref{fig:sb} in the $BVI$ bands. The galaxy is clearly not as extremely faint as some LSB galaxies identified as optical counterparts to dark galaxy candidates, but its central (i.e., maximum) surface brightness remains still just below the Pan-STARRS1 survey limit.

\subsection{Stellar mass estimation}
\label{subsection:stellarmass}

We derived the total stellar mass ($M_\star$) of the galaxy by combining our deep photometry with standard stellar population synthesis (SPS) scaling relations. First, we calculated the total $B$-band luminosity ($L_B$) by integrating the surface brightness profile (Fig.~\ref{fig:sb}, top panel). Adopting the HI-derived distance of $D = 28.8 \pm 2.9\,\mathrm{Mpc}$ \citep{Xu+2023ApJ...944L..40X}, we measured a total absolute luminosity of $L_B = (1.11 \pm 0.23)\times 10^{7}\,L_\odot$. The uncertainty includes both the uncertainty in the distance estimate and a conservative estimate of the photometric flux error. 

We converted the B-band luminosity into stellar mass by estimating the mass-to-light ratio ($M_\star/L_B$) from the galaxy’s photometrically measured color, using a color–M/L relation from SPS models.
The $(B-V)$ color profile (Fig.~\ref{fig:sb}, bottom panel) indicates a relatively blue stellar population in the inner region ($R \lesssim 9\, \mathrm{arcsec}$), clustering around a mean value of $(B-V) \simeq 0.48 \pm 0.18$. Stellar masses were estimated using the color–mass-to-light relation of \cite{IntoPo-SPS}, with a disk galaxy model and a Kroupa IMF, commonly adopted for low-mass galaxies \citep{Kroupa2001MNRAS.322..231K, Herrmann+2016AJ....152..177H}. This yields a mass-to-light ratio of $M_\star/L_B \simeq 0.65\,M_\odot/L_\odot$. Finally, this results in a total stellar mass of $M_\star = (7.2 \pm 3.7) \times 10^6\, M_{\odot}$. The uncertainty includes errors in distance, photometry, and the intrinsic scatter of the SPS relations. Although this mass estimate is lower than that derived for a passive red population, it remains significantly higher than the upper limit reported by \citet{Xu+2023ApJ...944L..40X}. This confirms the presence of a stellar counterpart, whereas the low stellar mass implies an extremely high gas fraction of $M_{\mathrm{HI}}/M_\star = 11.5 \pm 6.4$.

\subsection{Spectroscopic confirmation}\label{sec:spectra}

The spectrum of the galaxy shows a very faint stellar continuum and only a single emission line over the entire studied spectral range (see Fig.~\ref{fig:spectra}, and Appendix~\ref{sec:spec_discussion} for a more detailed discussion of the ionized gas properties). We interpret this feature as a redshifted H$\alpha$ emission line, as no alternative identification is plausible given the wavelength coverage, the absence of additional emission lines, and the angular extent of the source. The corresponding line-of-sight velocity in the integrated galactic spectrum ($\pm5''$ relative to the photometric center) is $V_{hel}=2492\pm25$~km~s$^{-1}$. Although the line is weak ($S/N\approx5$), the obtained velocity agrees with the FAST HI measurements within the $1.1\sigma$ interval, providing independent spectroscopic support for the physical association between the stellar and gaseous components. Future deep spectroscopic observations will allow us to extract additional information on the ionized gas and integrated stellar properties of the system.

\subsection{Reclassification as a gas-rich LSB dwarf galaxy}

Spectroscopic detection of H$\alpha$ emission at a velocity consistent with the HI source definitively establishes the physical association between the stellar and gaseous components. Consequently, we adopted the HI distance of $D=28.8\pm 2.9\,\mathrm{Mpc}$ for the galaxy. Their physical association, however, should also be clear even without this spectroscopic confirmation. The field is isolated with no other plausible optical candidates within the FAST beam width. The probability of a chance alignment between a massive, rotating HI cloud and a background LSB galaxy exactly at its peak column density is negligible. Furthermore, the morphology of the optical counterpart, an irregular, diffuse disk with a significantly elongated projected morphology, is consistent with the kinematic properties of the HI source described by \citet{Xu+2023ApJ...944L..40X}. This apparent elongation does not necessarily contradict the low inclination inferred from the HI data, which is affected by significant uncertainty due to the coarse spatial resolution of the FAST observations and the potential gas–stellar misalignment.

In Table~\ref{tab:properties}, we summarize the properties of FAST J0139+4328 derived from our deep imaging and spectroscopy. Despite its extreme gas richness, FAST J0139+4328 is not exceptionally unique. The derived $M_{\mathrm{HI}} / M_{\star}$ ratio, while high, is comparable to other gas-rich dwarfs found in the ALFALFA survey \citep[e.g.,][]{Bradford+2015ApJ...809..146B} and fits within the scatter of the stellar mass–gas fraction relation for LSB galaxies \citep[e.g.,][]{Du+2024ApJ...964...85D}, although it is one of the most gas-rich examples.

\begin{table}[ht]
\caption{Comparison of properties for FAST J0139+4328.}
\label{tab:properties}
\centering
\resizebox{\columnwidth}{!}{
\begin{tabular}{l c c}
\hline\hline
Parameter & \citet{Xu+2023ApJ...944L..40X} & This Work \\
\hline
\multicolumn{3}{c}{\textit{General Properties}} \\ \\
R.A. (J2000) & $01^{\mathrm{h}}39^{\mathrm{m}}28\fs4$ & $01^{\mathrm{h}}39^{\mathrm{m}}29\fs6$ \\
Dec. (J2000) & $+43\degr28\arcmin02\farcs6$ & $+43\degr28\arcmin29\farcs8$ \\
Distance $D$ [Mpc] & $28.8 \pm 2.9$ & $28.8 \pm 2.9$ (Adopted) \\
$V_{\mathrm{sys}}$ [km s$^{-1}$] & $2464.4 \pm 0.8$ (HI) & $2492 \pm 25$ (H$\alpha$) \\
\hline
\multicolumn{3}{c}{\textit{Derived Properties}} \\ \\
$M_{\mathrm{HI}}$ [$M_{\odot}$] & $(8.3 \pm 1.7) \times 10^7$ & -- \\
$L_B$ [$L_{\odot}$] & $< 1.4 \times 10^6$ & $(1.11 \pm 0.23)\times 10^{7}$ \\
$M_{\star}$ [$M_{\odot}$] & $< 6.9 \times 10^5$ & $(7.2 \pm 3.7) \times 10^6$ \\
$M_{\mathrm{HI}} / M_{\star}$ & $> 120$ & $11.5 \pm 6.4$ \\
Classification & Dark Galaxy Candidate & Gas-Rich LSB Dwarf \\
\hline
\end{tabular}}
\tablefoot{
Values from \citet{Xu+2023ApJ...944L..40X} represent the HI properties and optical upper limits from Pan-STARRS1.
Our values represent the properties of the optical counterpart derived from deep $LBVI$ imaging and H$\alpha$ spectroscopy.
}
\end{table}

\section{Conclusion}\label{sec:conclusion}

We presented the discovery of a faint optical counterpart to the HI source FAST J0139+4328, previously claimed to be the first isolated dark galaxy in the local universe. Using deep imaging from the 1.4m \textit{Milanković} telescope and the 0.6m \textit{Nedeljković} telescope, we revealed a stellar component that was missed by the Pan-STARRS1 survey.

Our analysis classifies FAST J0139+4328 as a gas-rich LSB dwarf galaxy. We derive a stellar mass of $M_\star = (7.2 \pm 3.7) \times 10^6\, M_{\odot}$, about an order of magnitude higher than the previously estimated upper limit \citep{Xu+2023ApJ...944L..40X}. Using the literature HI mass, this implies a gas-to-stellar mass ratio of $M_{\mathrm{HI}} / M_{\star} = 11.5 \pm 6.4$, placing it among the gas-rich tail of the dwarf galaxy population. Crucially, long-slit spectroscopy confirmed the physical association between the stellar body and the HI cloud by detecting H$\alpha$ emission at the expected redshift.

We therefore conclude that FAST J0139+4328 is not a dark galaxy but rather an extremely gas-rich LSB dwarf galaxy. This reclassification resolves the status of this prominent dark galaxy candidate and highlights that other objects identified by HI surveys may be LSB systems awaiting deeper optical detection. It underscores that non-detections in medium-depth surveys are insufficient to rule out the presence of stellar populations in faint HI systems. Future studies of similar candidates require deep optical imaging to distinguish between truly starless halos and the extreme LSB galaxy population. We plan further deep spectroscopic observations to obtain detailed insights into the nature, characteristics, and evolutionary pathway of FAST J0139+4328.

\begin{acknowledgements}
We thank the staff of the Astronomical Station Vidojevica for their excellent support and assistance during the observing runs. This research was supported by the Ministry of Science, Technological Development and Innovation of the Republic of Serbia (MSTDIRS) through contract no. 451-03-136/2025-03/200002, made with the Astronomical Observatory (Belgrade, Serbia), and contract no. 451-03-136/2025-03/200104 made with the Faculty of Mathematics, University of Belgrade. Observations with the SAO RAS telescopes are supported by the Ministry of Science and Higher Education of the Russian Federation. The renovation of telescope equipment is currently provided within the national project ``Science and Universities''. The work on optical spectroscopy was performed as part of the SAO RAS government contract approved by the Ministry of Science and Higher Education of the Russian Federation.
\end{acknowledgements}

\bibliographystyle{aa} 
\bibliography{aa58391-25}

\begin{appendix} 
\section{Spectroscopic discussion}\label{sec:spec_discussion}

\begin{figure}[ht]
    \centering
    \includegraphics[width=1.\linewidth]{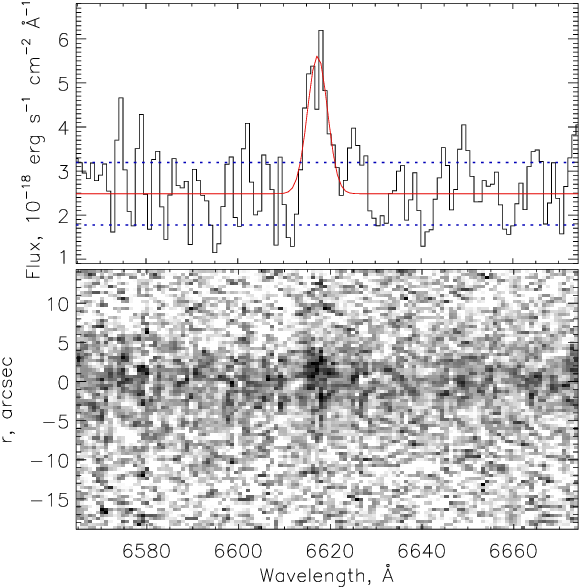}
    \caption{Fragment of the 2D spectrum obtained with BTA/SCORPIO-2  (bottom) and integrated spectrum (10\farcs1 along the slit) with Gaussian fitting of the H$\alpha$ emission line (top). The blue dotted lines indicate $\pm 1 \sigma$ around the stellar continuum level.}
    \label{fig:spectra}
\end{figure}

Fig.~\ref{fig:spectra} shows the optical spectrum of FAST J0139+4328 obtained as described in Sect.~\ref{sec:obs}, intending to independently verify the association between the stellar component and the HI cloud. The spectrum shows a faint redshifted H$\alpha$ emission line (see Sect.~\ref{sec:spectra} for a justification) at a velocity consistent with the HI measurement.

The flux of the integrated emission line $F(H\alpha)=1.6\cdot10^{-17}$~erg~s$^{-1}$~cm$^{-2}$ corresponds to the mean surface brightness $log(\Sigma_{H\alpha})=37.0$~erg~s$^{-1}$~kpc$^{-2}$. This value is significantly lower than the upper limit usually accepted for the selection of diffuse ionizing gas (DIG) in galaxies \citep[$log(\Sigma_{H\alpha})=39.0$,][]{Zhang2017MNRAS.466.3217Z}. There are no forbidden emission lines and bright H$\alpha$ clumps at our detection level. This can hint at the possibility that observed DIG emission is related to hot low-mass evolved stars ionizing the interstellar medium. New deep spectral observations are needed to get a more detailed conclusion about the sources of gas ionization in FAST J0139+4328.

\end{appendix}

\end{document}